# Long Range Antiferromagnetic Ordering in the $S$ = ½ Ordered Rock Salt Oxide, $Li_5OsO_6$: Comparison with the Isoelectronic and Isostructural Spin Glass, $Li_4MgReO_6$

Shahab Derakhshan, [1, 2, *], John E. Greedan [1, 2]

[1,2] Brockhouse Institute for Materials Research and Chemistry Department, McMaster University, 1280 Main Street West, Hamilton, Ontario, Canada, L8S 4M1.

and

Lachlan M. D. Cranswick [3]

[3] Canadian Neutron Beam Centre, National Research Council Canada Building 459, Chalk River Laboratories, Chalk River ON, Canada, K0J 1J0

*: To whom correspondence should be addressed. Email: derakh@mcmaster.ca

**Abstract.** $Li_5OsO_6$ and $Li_5ReO_6$ polycrystalline samples were synthesized by conventional solid state methods. Employing powder neutron diffraction data, the crystal structure of $Li_5OsO_6$ was re-investigated. $Li_5OsO_6$ crystallizes in $C2/m$ space group in an ordered NaCl structure type where $a$ = 5.0472(1) Å, $b$ = 8.7827(2) Å, $c$ = 5.0079(1) Å, $\beta$ = 109.777(2)° and $V$ = 208.90(1) Å$^3$. Magnetic susceptibility and heat capacity data indicate an AFM long range order below 40K although there is evidence for low dimensional short range order above 80K. As well, the frustration index, f = $|\theta|/T_N$ ~1, in contrast to the isostructural and isoelectronic compound, $Li_4MgReO_6$, which is a spin glass below 12K and has f ~ 14. An attempt was made to rationalize these differences using spin dimer analysis. The key results are that the spin exchange interactions in the Re-based compound are stronger and are consistent with a frustrated triangular lattice model while a low dimensional short range order model is better for $Li_5OsO_6$. The main reason for this is a strong *Jahn-Teller* distortion in the $OsO_6$ octahedron material which constrains the unpaired electron to occupy the $d_{xy}$ orbital.

PACS number(S):. 75.50.Ee, 75.50.Lk, 75.40.-s, 61.05.F-.

## INTRODUCTION

Antiferromagnetic (AFM) compounds with a triangular cationic sub-lattice have potential to exhibit magnetic frustration, as the three exchange interactions can not be satisfied simultaneously. [1] Such systems with $S = ½$ have been considered as spin liquid ground state candidates. [2]

Ordered rock salt type transition metal oxides [3] comprise such lattices and are interesting from this point of view. $LiNiO_2$, [4,5] and $NaTiO_2$ [6] have been considered as candidates for a spin-liquid magnetic ground state. While $Na_3Co_2SbO_6$ [7] with a honeycomb lattice exhibits AFM long range order, isostructural $Na_3Cu_2SbO_6$ [8] has a singlet ground state with spin gap behavior and is best defined based on a short-range AF-AF one dimensional alternate chain model. [9]

In contrast to the *3d* transition metal oxides, the physical properties of *4d* and *5d* compounds have not been widely studied. The latter compounds have more extended *d* orbitals and also exhibit large spin-orbit coupling. Accordingly, they usually behave differently from the 3*d* systems. $Na_3RuO_4$ [10] exhibits three-dimensional magnetic ordering with some degrees of frustration. Furthermore, $Li_4MgReO_6$ [11] shows spin-glass behavior below 12K and not long range order. The chemistry of osmium oxides is very rich as Os can take a wide range of valence numbers, namely +4 - +7. Among the $Os^{7+}$ compounds, $Ba_2LiOsO_6$ undergoes a long range AFM ordering below 8K with some evidence of frustration whereas $Ba_2NaOsO_6$ shows FM behavior.[12, 13]

The physical properties of the title compound, $Li_5OsO_6$ [14], which is isostructural and isoelectronic with $Li_4MgReO_6$, have not been investigated in detail. The existing, but sparse, magnetic susceptibility data indicate a possible anomaly near 40K. In this article the temperature dependent magnetic susceptibility and heat capacity data for $Li_5OsO_6$ are measured and are compared to those of $Li_4MgReO_6$. Moreover, to understand the driving force for the different magnetic properties between two systems, the calculated relative magnitudes of the different spin exchange interactions for both compounds are also presented.

## II. EXPERIMENTAL SECTION

### A. Synthesis

The starting material, $Li_2O$ was prepared by heating lithium hydroxide mono-hydrate (Frederick Smith Chemicals, 98%) to 450°C in a fused silica tube under dynamic vacuum for 18 hours. The $Li_5OsO_6$ sample was prepared according to the procedure introduced by Betz *et al.* [14] For this purpose a stoichiometric mixture of $Li_2O$ and Os (Alfa Aesar, 99.95%) powder was thoroughly ground and pressed into a pellet in an argon filled glove box. The pellet was placed in an alumina boat and heated to 500 °C in a tube furnace under dynamic argon flow, after five hours the argon flow was switched to oxygen flow and the sample was heated to 800 °C for twelve hours and was cooled to room temperature in ten hours.

For preparing $Li_5ReO_6$, a stoichiometric mixture of $Li_2O$ and $Re_2O_7$ (CERAC, 99.99% ) powder was pressed into a pellet, which was placed in a platinum boat and heated to 900 °C in a tube furnace under dynamic oxygen flow. After 18 hours the furnace was turned off. [15]

### B. Phase Analyses

The formation and phase purity of the black $Li_5OsO_6$ and yellow $Li_5ReO_6$ products were confirmed using powder X-ray diffraction, employing a Guinier-Hägg camera with Cu $K\alpha_1$ radiation and Si as the internal standard. To convert the film record to digital data a KEJ line scanner was utilized.

### C. Crystal Structure and Magnetic Structure Determination Using Neutron Diffraction

Powder neutron diffraction measurements at different temperatures were performed on the C2 diffractometer at the Canadian Neutron Beam Centre at Chalk River, Ontario. The room temperature data were collected using two different wavelengths of 1.3307 Å in the angular range of $36^o \leq 2\theta \leq 113^o$ with $0.1^o$ steps and 2.3724 Å in the angular range of $10^o \leq 2\theta \leq 82^o$.

Low temperature data (3K) were collected for investigation of the magnetic structure. For this purpose long wavelength of 2.36957 Å in the range $5^o \leq 2\theta \leq 82^o$ with $0.1^o$ interval was utilized.

### D. Physical Properties Measurements

Temperature dependent magnetic susceptibility data for a $Li_5OsO_6$ powder sample, encased in a gelatin capsule, were collected employing a Quantum Design MPMS SQUID magnetometer. Both zero-field cooled (ZFC) and field cooled (FC) data were obtained over the temperature range of 5-300 K at an applied field of 1000 Oe. Diamagnetic corrections were added to the susceptibility data.

Heat capacity data were collected from 5-60 K using the heat capacity probe of the Oxford MagLab system, without applying any external magnetic field. The powder sample was pressed into a thin pellet and a small portion of that was re-sintered to minimize the grain boundaries. The thin block was mounted onto a sapphire measurement chip with Apeizon grease. Contributions to the measured heat capacity by the grease and sample holder chip were calibrated.

## E. Theoretical Calculations, Spin Dimer Analyses

To estimate the relative values of the various exchange constants, *J*'s, extended Hückel, spin dimer analysis [16] was performed. In these computations, two $OsO_6^{5-}$ units ($Os_2O_{12}^{10-}$ dimer) for each pathway were taken into account and the inter site hopping energy, ($\Delta e$), was estimated using the *CAESAR* package. [17] Double-zeta Slater Type Orbitals (STO) were employed for the oxygen *s* and *p* and osmium *d* states and single-zeta for osmium *s* and *p* states. The values of the $\zeta_i$ and $\zeta'_i$ coefficients and valence shell ionization potentials $H_{ii}$ used for the calculations are presented in Table I.

For $d^1$ systems the tetragonal compression *Jahn-Teller* effect is expected, where the $d_{xy}$ state lies below the $d_{xz}$ and $d_{yz}$ states. In extreme conditions the $d_{xy}$ state is well separated from the other states and exclusively accommodates the unpaired electron. In this case there is only one interaction, which is responsible for the magnetic interaction and assuming that $J \cong \frac{(\Delta e)^2}{U}$ and that *U* is constant, the relative magnitude of the various *J*'s can be determined.

Alternatively, if the energy separations between these states are small, the probability of the $d_{xz}$ and $d_{yz}$ states to be occupied will be high and they can also contribute to the spin exchange interactions. In the simplest case the probability that $d_{xy}$, $d_{xz}$ and $d_{yz}$ contribute to the exchange interactions is equal and we have:

$$< (\Delta \mathrm{e})^2 > \approx \frac{1}{N^2}\sum_{\mu=1}^{N}(\Delta e_{\mu\mu})^2 \qquad (1)$$

In these systems there are three states which are involved and therefore the equation (1) can be re arranged as:

$$< (\Delta \mathrm{e})^2 > \approx \frac{1}{9}[(\Delta e_{11})^2 + (\Delta e_{22})^2 + (\Delta e_{33})^2] \qquad (2)$$

and finally the spin exchange interactions will be given by:

$$J \cong \frac{<(\Delta e)^2>}{U} \qquad (3)$$

The corresponding relative exchange interactions for both Re and Os-based compounds were calculated according to the both abovementioned approaches. To examine the consistency of the calculations, different values of (1-x)× $\zeta'_i$ with x =0, 0.05 and 0.1 for oxygen 2*p* atomic orbitals were employed. $\zeta'_i$ describes the diffuse STO and by providing an orbital tail, enhances the overlap between oxygen atoms within the SSE pathways. [18] To compare the relative strengths of the *Jahn-Teller* effect for these two systems, extended Hückel molecular orbital calculations for $[OsO_6]^{5-}$ and $[ReO_6]^{6-}$ species were performed as well.

**Table I.** The values for the $\zeta_i$ coefficients and valence shell ionization potentials $H_{ii}$ of the STO's employed for the spin dimer calculations for $Li_5OsO_6$ and $Li_4MgReO_6$.

| Atom | Orbital | $H_{ii}$ (eV) | $\zeta_i$ | C | $\zeta'_i$ | C' |
|---|---|---|---|---|---|---|
| O | *2s* | -32.300 | 2.688 | 0.7076 | 1.675 | 0.3745 |
| O | *2p* | -14.8000 | 3.694 | 0.3322 | 1.659[a] | 0.7448 |
| Os | *6s* | -8.170 | 2.400 | 1 | | |
| Os | *6p* | -4.810 | 1.770 | 1 | | |
| Os | *5d* | -11.840 | 4.504 | 0.6066 | 2.391 | 0.5486 |
| Re | *6s* | -9.360 | 2.346 | 1 | - | - |
| Re | *6p* | -5.960 | 1.730 | 1 | - | - |
| Re | *5d* | -12.660 | 4.339 | 0.5886 | 2.309 | 0.5627 |

a: This value corresponds to the diffuse STO of O *2p* when x = 0.

## III. RESULTS AND DISCUSSION

### A. Crystal structure

$Li_5OsO_6$ crystallizes in *C*2/*m* in an ordered form of the rock salt structure type (Figure 1). The structure is composed of edge shared octahedra where one layer has the composition of $Li_2OsO_6^{3-}$ and the other layer only contains $Li^+$ cations ($LiO_6$ octahedra).

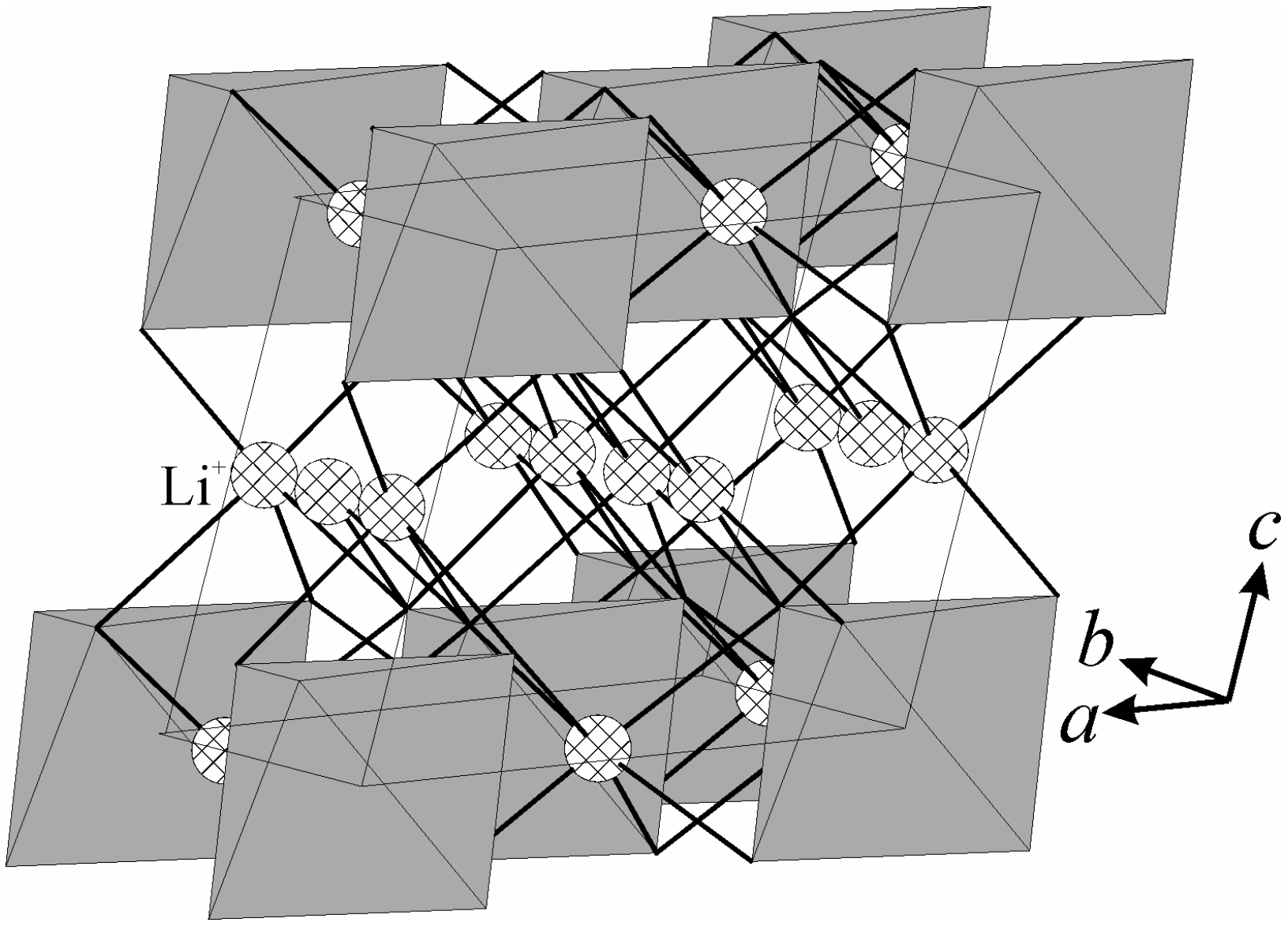

**Figure 1:** Edge-sharing octahedra in $Li_5OsO_6$ viewed. The grey octahedra represent $[OsO_6]^{5-}$ and the circles are $Li^+$ ions.

To investigate the crystal structure, Rietveld refinements were performed on the room temperature neutron diffraction data sets using the GSAS program.[19, 20] The cell parameters and atomic positions were initially taken from the proposed model by Betz *et al.*[14]. A pseudo-Voigt peak shape profile, which is composed of both Gaussian and Lorentzian parameters, was chosen and the parameters were refined to obtain the best fit to the experimental data. The space group is *C*2/*m*, with lattice dimensions of $a$ = 5.0472(1) Å, $b$ = 8.7827(2) Å, $c$ = 5.0079(1) Å, $\beta$ =

109.777(2)°, $V$ = 208.90(1) $Å^3$, with residual factors of Rp = 0.032 and wRp = 0.042 (Figure 2). $Li_5OsO_6$ is isostructural to $Li_4MgReO_6$. [11] In addition to the fact that $Li_5OsO_6$ has a smaller cell volume than $Li_4MgReO_6$ the major structural difference is that the three independent crystallographic positions for Li in $Li_5OsO_6$ exhibit mixing between Li and Mg in $Li_4MgReO_6$. The crystallographic details and the atomic positions are summarized in Tables II and III. Some selected interatomic distances are presented in table IV. In both $Li_5OsO_6$ and $Li_4MgReO_6$ the $Os(Re)O_6$ octahedra exhibit a tetragonal compression, with two bonds shortened, which may be due to a *Jahn-Teller* distortion which is expected for a $d^1$ electronic configuration. This feature is evidently more pronounced in $Li_5OsO_6$.

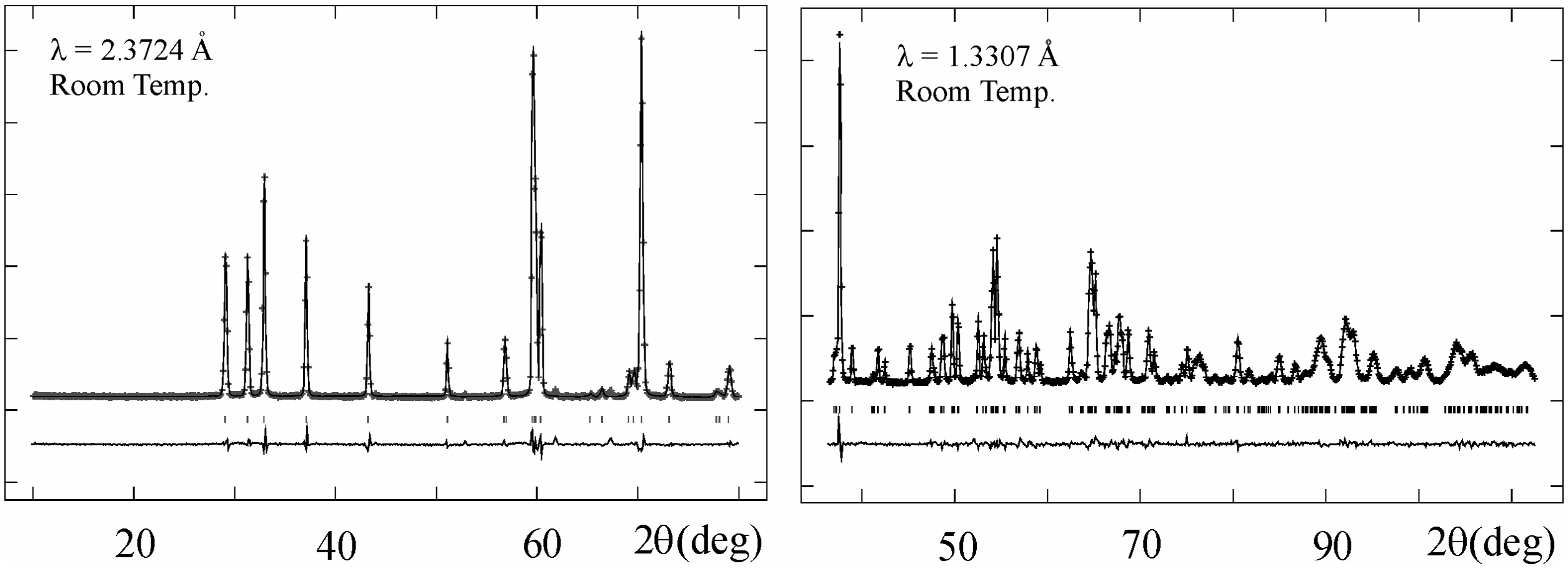

**Figure 2:** Room temperature neutron diffraction pattern using long wavelength (left) and short wavelength (right) neutrons. The cross points indicate the experimental data, the solid line represents the Rietveld fit and the thin lines bellow the pattern the difference. The expected peak positions are shown by the tick marks.

**Table II.** Some selected structural parameters of $Li_5OsO_6$ and $Li_4MgReO_6$.

| | $Li_5OsO_6$ | $Li_4MgReO_6$ [11] |
|---|---|---|
| $a$ (Å) | 5.0472(1) | 5.0979 |
| $b$ (Å) | 8.7827(2) | 8.8163 |
| $c$ (Å) | 5.0079(1) | 5.0815 |
| $\beta$ (°) | 109.777(2) | 109.835 |
| $V$ (Å$^3$) | 208.90(1) | 214.83 |

**Table III.** Atomic coordinates and equivalent isotropic displacement parameters for $Li_5OsO_6$

| | $X$ | $Y$ | $Z$ | $U_{iso}$ (Å$^2$) |
|---|---|---|---|---|
| Os | 0 | 0 | 0 | 0.0057(3) |
| Li1 | 0 | 0.6733(8) | 0 | 0.013(1) |
| Li2 | 0 | 0.5 | 0.5 | 0.014(2) |
| Li3 | 0.5 | 0.3148(7) | 0.5 | 0.014(2) |
| O1 | 0.2688(4) | 0.3467(2) | 0.7634(3) | 0.0068(4) |
| O2 | 0.2741(5) | 0.5 | 0.2229(4) | 0.0099(6) |

**Table IV.** Some selected interatomic Os–O and Re–O distances of $Li_5OsO_6$ and $Li_4MgReO_6$.

| $Li_5OsO_6$ | | $Li_4MgReO_6$ [11] | |
|---|---|---|---|
| 4 × Os–O (Å) | 1.9083(1) | 4 × Re–O (Å) | 1.9622 |
| 2 × Os–O (Å) | 1.8459(2) | 2 × Re–O (Å) | 1.9323 |

## B. Magnetic Susceptibilities

In the low temperature data (5–100K) the sharp increase in susceptibility at low temperatures (5-20 K) regime (Figure 3) can be attributed to a paramagnetic impurity. A sharp lambda-shaped peak in the susceptibility with a maximum near 40 K, is indicative of the long range antiferromagnetic nature of the magnetism of this compound.

The high temperature magnetic susceptibility data of $Li_5OsO_6$ from 100-300 K (Figure 3. b) were seen to fit very well to the Curie-Weiss law, $\chi = \frac{C}{T-\theta}$ . The fitting parameters are $C$ = 0.1068(2) emu/mol for the Curie constant and $\theta$ = -34.0(5) K for the Weiss constant. This Curie constant corresponds to an effective magnetic moment, $\mu_{eff}$, of 0.92(4) $\mu_B$ per $Os^{7+}$ ($5d^1$, $S$ = ½), which is lower than the spin only value of 1.73 $\mu_B$. This is in agreement with the fact that the magnetic moment of the heavy cations, such as $Os^{7+}$, is strongly influenced by orbital contributions as well, and spin-orbit coupling in the electronic configurations less than half filled is in the form of L – S. This results in a magnetic moment with a lower value than that of the spin-only case. The negative value for the Weiss constant is indicative of predominant AF interactions.

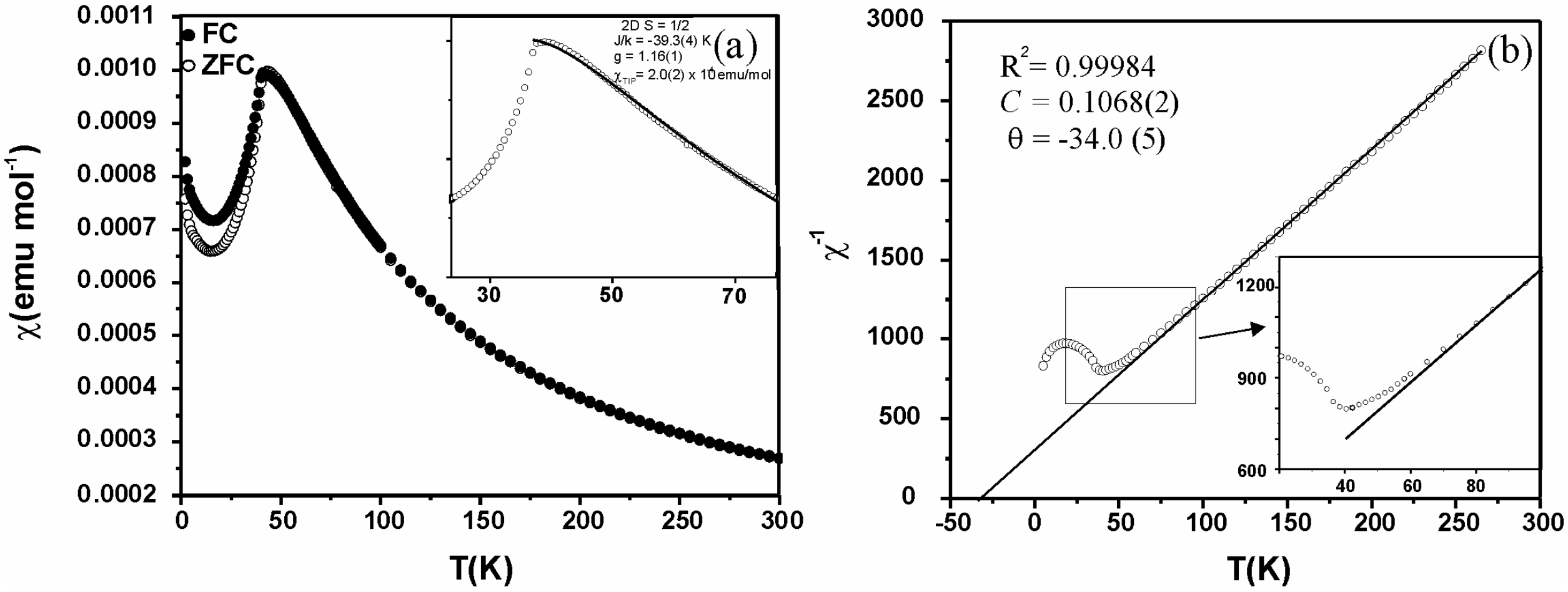

**Figure 3:** (a) Zero-field cooled and field cooled magnetic susceptibility data for $Li_5OsO_6$. The open circles represent the ZFC data and the filled circles represent the FC data. (b) Curie Weiss fit in the high temperature, paramagnetic, region. The open circles represent the ZFC data points and the solid line is the fit. The insets show evidence for short range correlations which are described in the text.

The relationship between the Weiss temperature and the various exchange constants, $J_m$, is well-known and given as equation 1 by the Mean Field Theory: [21]

$$\theta = \frac{2S(S+1)}{3k_B}\sum_{m=1}^{N} z_m J_m \qquad (4)$$

where $\theta$ is the Weiss constant, $z_m$ is the number of $m^{th}$ nearest neighbors of a given atom, $J_m$ is the exchange interaction between $m^{th}$ neighbors and $N$ is the number of sets of neighbors for which $J_m \neq 0$. The Weiss temperature sets the scale for the magnetic exchange energy in a system. In general, large exchange energies result in relatively high Neél temperatures, unless some other factors like frustration prevents the ordering and lowers the transition temperature. It is common to apply the so-called frustration index, f ~ $\frac{\theta}{T_N}$ to determine the level of magnetic frustration in a material. [22] Systems showing $f > 5$ are considered frustrated and therefore $Li_5OsO_6$ with $f \sim 1$ is not a frustrated system. As will be discussed later, this is in stark contrast to the situation for $Li_4MgReO_6$.

C. Heat capacity data

The temperature dependent heat capacity data of $Li_5OsO_6$ is shown in Figure 4(a) as black circles. A lambda-shaped anomaly is seen near 40K, which is additional proof for the long range magnetic order at this temperature. Heat capacity is composed of both electronic and phonon (lattice) contributions. To eliminate the lattice component we collected comparable data for $Li_5ReO_6$ (open circles in Figure 4.a). This material was the best choice for this purpose as: (1) It crystallizes in the same structure type, (2) The electronic configuration of $Re^{7+}$ is *[Xe]* $5d^0$ $6s^0$ and there is no unpaired electron in the outer shell to contribute in heat capacity, (3) Re is adjacent to Os in periodic table and the mass difference is minimized.

Subtracting these data sets, the electronic contribution to the heat capacity of $Li_5OsO_6$ was obtained (Figure 4.b). Note the remarkable accordance with the Fisher heat capacity analysis [23] (the inset in Figure 4.b) using magnetic susceptibility data. For this purpose the derivative of the $\chi$T to T ($\frac{d(\chi.T)}{dT}$) is plotted versus T, which gives a good approximation to the magnetic component of the heat capacity.

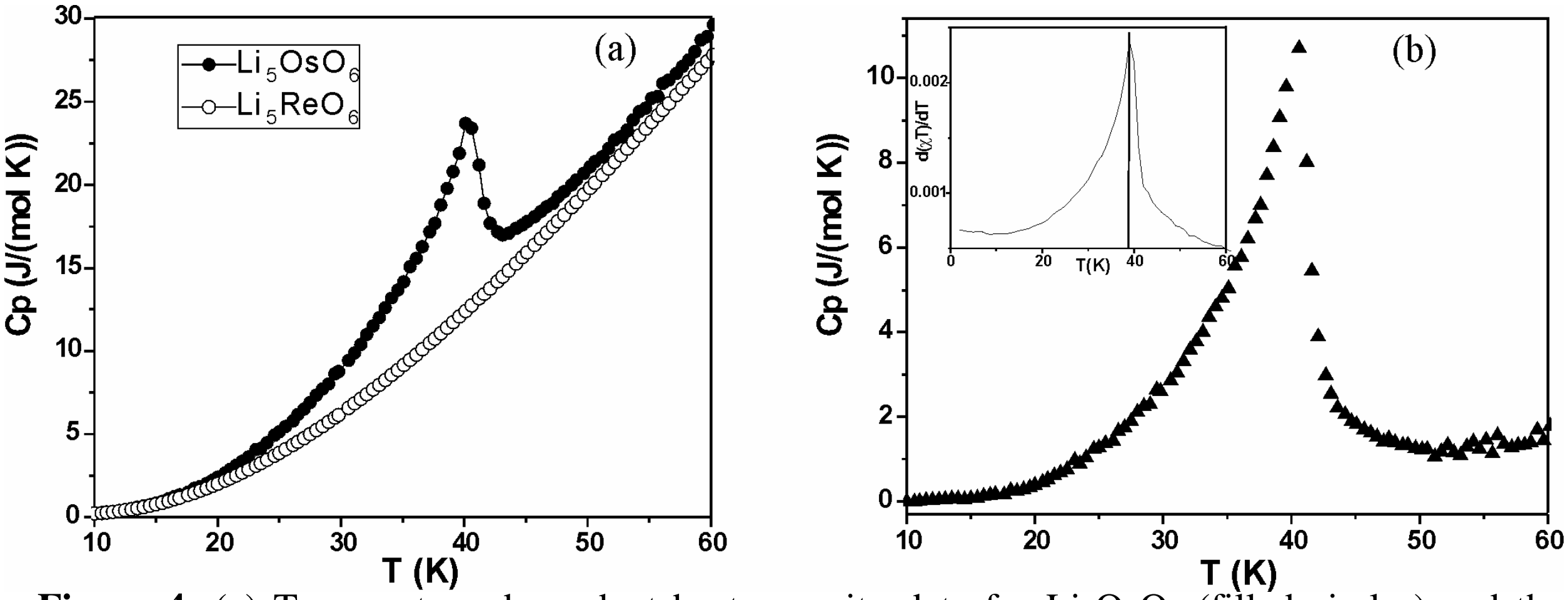

**Figure 4:** (a) Temperature dependent heat capacity data for $Li_5OsO_6$ (filled circles) and the lattice match $Li_5ReO_6$ (open circles). (b) The difference plot represents the electronic contribution to the heat capacity, which is in a very good agreement with Fisher heat capacity analyses (inset plot) using magnetic data.

One can calculate the entropy associated with the transition using equation 2:

$$\Delta S_{\exp} = \int_0^T \left(\frac{C_P}{T}\right) dT \qquad (5)$$

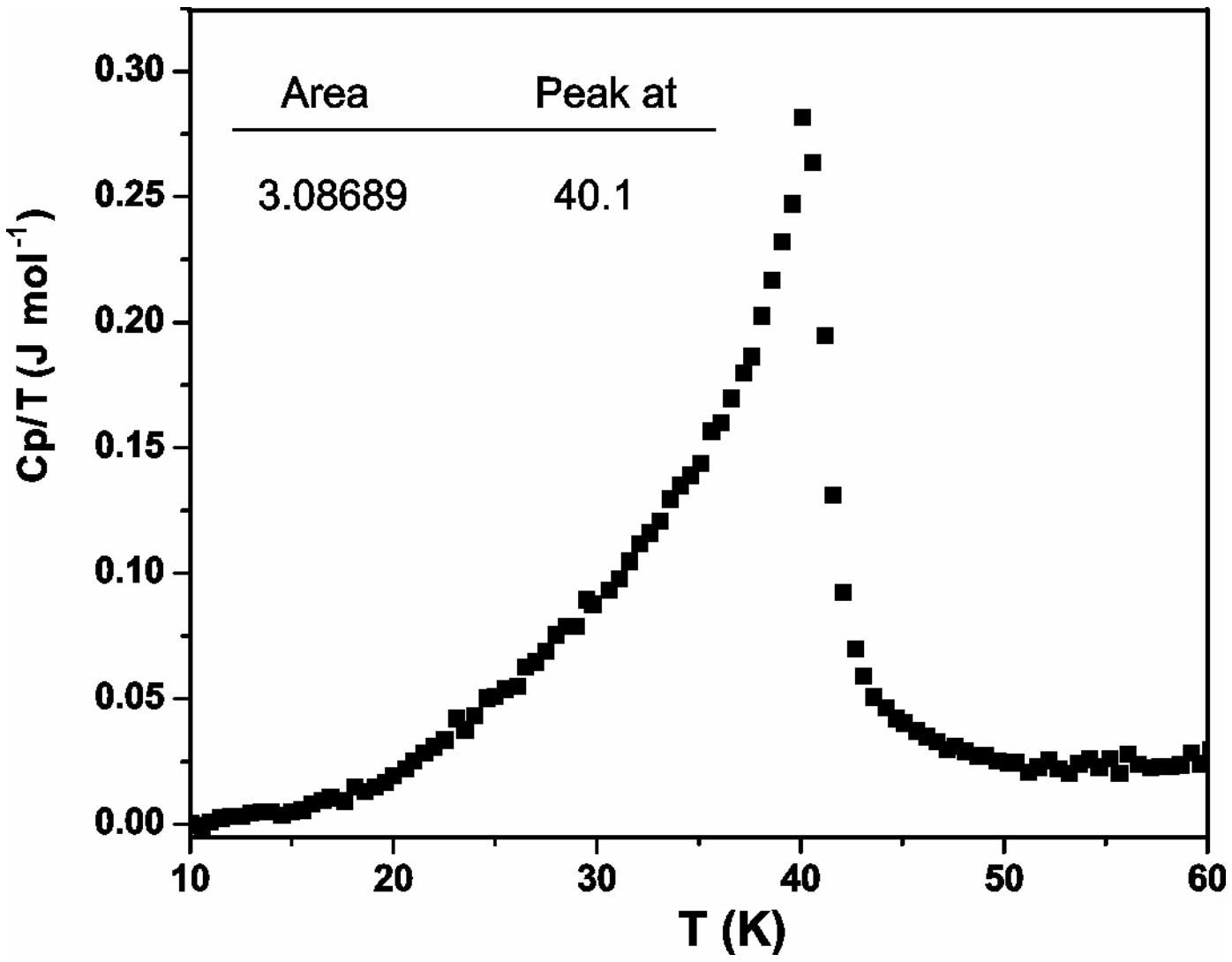

**Figure 5:** $\frac{C_P}{T}$ as the function of the temperature. The area under the peak represents the entropy for the magnetic transition in $Li_5OsO_6$.

Accordingly, the peak area in Figure 5 is the experimental entropy associated with the magnetic phase transition within the temperature range, which is 3.09 J/ (mol.K). On the other hand, Boltzmann's law gives the overall theoretical transition entropy:

$$\Delta S_{theor} = R \times ln\ \omega \qquad (6)$$

where, $\omega$ is spin multiplicity; $2S + 1 = 2$ and therefore:

$$\Delta S_{theor} = 5.76 \text{ J/ (mol.K)}$$

$$\frac{\Delta S_{\exp}}{\Delta S_{theor}} = \frac{3.09}{5.76} = 0.54$$

So, only 54% of the total transition entropy is lost below the magnetic phase transition which suggests the importance of some short range interactions at higher temperatures.

Corroborating evidence for short range magnetic correlations can be found from Figure 3. Note (inset of Figure 3b) that deviations from the Curie Weiss law set in below ~80K, a factor of two greater than $T_N$. As well (Figure 3a) the susceptibility just above $T_N$ is convex upward, a typical signature of short range AFM correlations. The data within the interval 40K – 80K could be fitted to various $S$ = ½ 1D or 2D models (a 2D fit is shown in the inset to 3a) giving roughly equal residuals and $J_{SRO}/k_B$ ≈ -40K (2D) or ≈ -30K (1D). While this observation is strong support for the importance of antiferromagnetic SRO in $Li_5OsO_6$, it is not possible to assign a specific model given the relatively narrow temperature range and the absence of a well-defined susceptibility maximum. The computational results to be described in section E support a low dimensional model for the short range correlations rather than a geometrically frustrated model.

D. Comparison of $Li_5OsO_6$ with $Li_4MgReO_6$

Given that $Li_5OsO_6$ is isostructural and isoelectronic with the previously studied $Li_4MgReO_6$ and both are $S$ = ½ spin systems, a detailed comparison is in order. Table V shows a comparison of parameters relating to the issues of frustration and the nature of the magnetic ground state.

**Table V.** Comparison of relevant magnetic parameters for $Li_4MgReO_6$ and $Li_5OsO_6$.

| | $Li_4MgReO_6$ | $Li_5OsO_6$ |
|---|---|---|
| $\mu_{eff}$ ($\mu_B$) | 1.14 | 0.92(4) |
| $\theta$(K) | -166(3) | -34.0(5) |
| $T_{N,g}$ | ~12 | 40 |
| Ground state | spin glass | AF LRO |
| $F$ | ~14 | ~1 |

While there is one similarity, the effective magnetic moment is strongly reduced for both materials from the spin only value of 1.73 $\mu_B$, due to the effects of spin-orbit coupling, in most cases the two materials are profoundly different. $Li_4MgReO_6$ is spin glass like below 12K with a very large, negative $\theta$ = -166K and a high frustration index of ~ 14. The Os-based material is on the other hand a more or less conventional antiferromagnet with $T_N$ ~ 40K and a frustration index near unity. This raises questions as to what are the factors which cause such a remarkable contrast in the properties of two isoelectronic and isostructural compounds.

## E. Computational methods

### *1. The tight binding, magnetic dimer model.*

All the possible spin exchange interaction pathways are presented in Figure 6. All these pathways are super-super-exchange (SSE) and their lengths and angles are summarized in Table VI and VII, respectively.

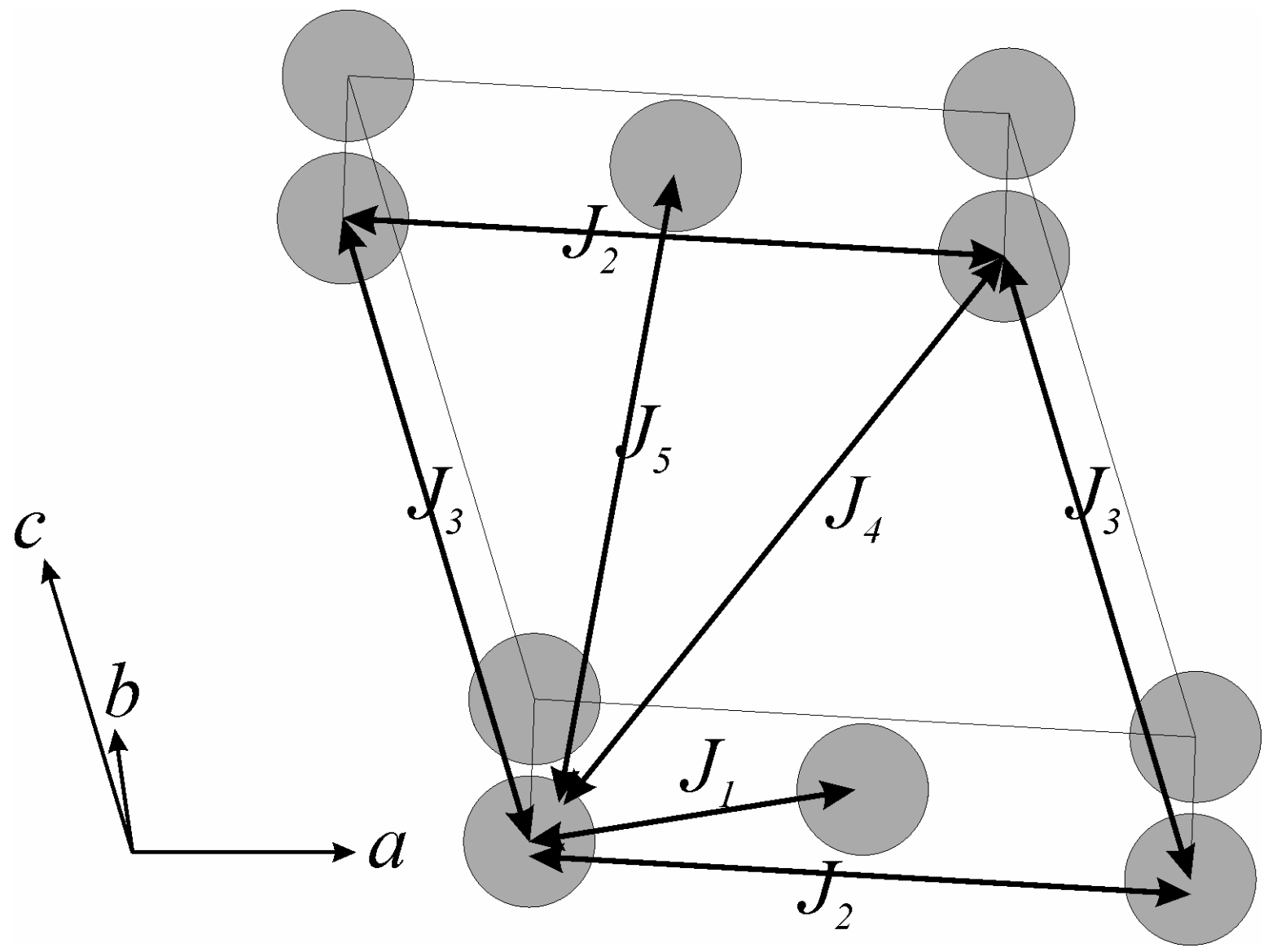

**Figure 6:** Schematic representation of different possible Os–Os interaction path ways in the unit cell.

**Table VI.** The relevant distances to the five identified exchange pathways, $J_1$, $J_2$, $J_3$, $J_4$ and $J_5$.

| | $M$ = Os | $M$ = Re [11] |
|---|---|---|
| ($M$–$M$)[$J_1$] (Å) | 5.065 | 5.0920 |
| ($M$–$M$)[$J_2$] (Å) | 5.047 | 5.098 |
| ($M$–$M$)[$J_3$] (Å) | 5.008 | 5.082 |
| ($M$–$M$)[$J_4$] (Å) | 5.783 | 5.851 |
| ($M$–$M$)[$J_5$] (Å) | 6.495 | 6.554 |

**Table VII.** The relevant ($M$ – O – O) angles for to the five identified exchange pathways, $J_1$, $J_2$, $J_3$, $J_4$ and $J_5$.

| | $M$ = Os | $M$ = Re [11] |
|---|---|---|
| (<$M$ – O – O)[$J_1$] (°) | 136.63 | 136.75 |
| (<$M$ – O – O)[$J_2$] (°) | 91.49 | 96.51 |
| (<$M$ – O – O)[$J_3$] (°) | 92.19 | 90.95 |
| (<$M$ – O – O)[$J_4$] (°) | 135.11 | 135.68 |
| (<$M$ – O – O)[$J_5$] (°) | 140.01 | 137.57 |

In the case of extreme *Jahn-Teller* distortion, the spin dimer analysis suggests that the $J_4$ interaction (along the 101 direction) is the largest interaction for both compounds and $J_1$ and $J_5$ with two orders of magnitude lower relative strength are the interactions which mediate the strong $J_4$ interaction in a 3D magnetic ordering. $J_2$ and $J_3$ are negligible for both systems.

**Table VIII.** $(\Delta e)^2$ for the various exchange pathways in both $Li_5OsO_6$ and $Li_4MgReO_6$ calculated on the spin dimer model assuming $d_{xy}$ occupation only.

| Pathway | $Li_5OsO_6$ $(\Delta e)^2$ $(meV)^2$ | Rel. | $Li_4MgReO_6$ $(\Delta e)^2$ $(meV)^2$ | Rel. |
|---|---|---|---|---|
| $J_1$ | 68.91 | **0.022** | 238.08 | **0.049** |
| $J_2$ | 0.14 | 0.00005 | 0.18 | 0.00004 |
| $J_3$ | 0.04 | 0.00001 | 1.32 | 0.0003 |
| $J_4$ | 3102.82 | **1** | 4854.74 | **1** |
| $J_5$ | 57.76 | **0.019** | 74.29 | **0.015** |

The fact that the $J_4$ is the strongest interaction, suggests that the interactions become more significant if the orbitals that accommodate the unpaired electrons are co-planar (Fig. 7). This approach results in a rather low dimensional magnetic structure, where the relative magnitudes of exchange interactions, $J_4 >> J_1$ and $J_5$, for both compounds do not support geometrical magnetic frustration. This is in agreement with the experimental magnetic data for the $Li_5OsO_6$, but not with those of $Li_4ReMgO_6$.

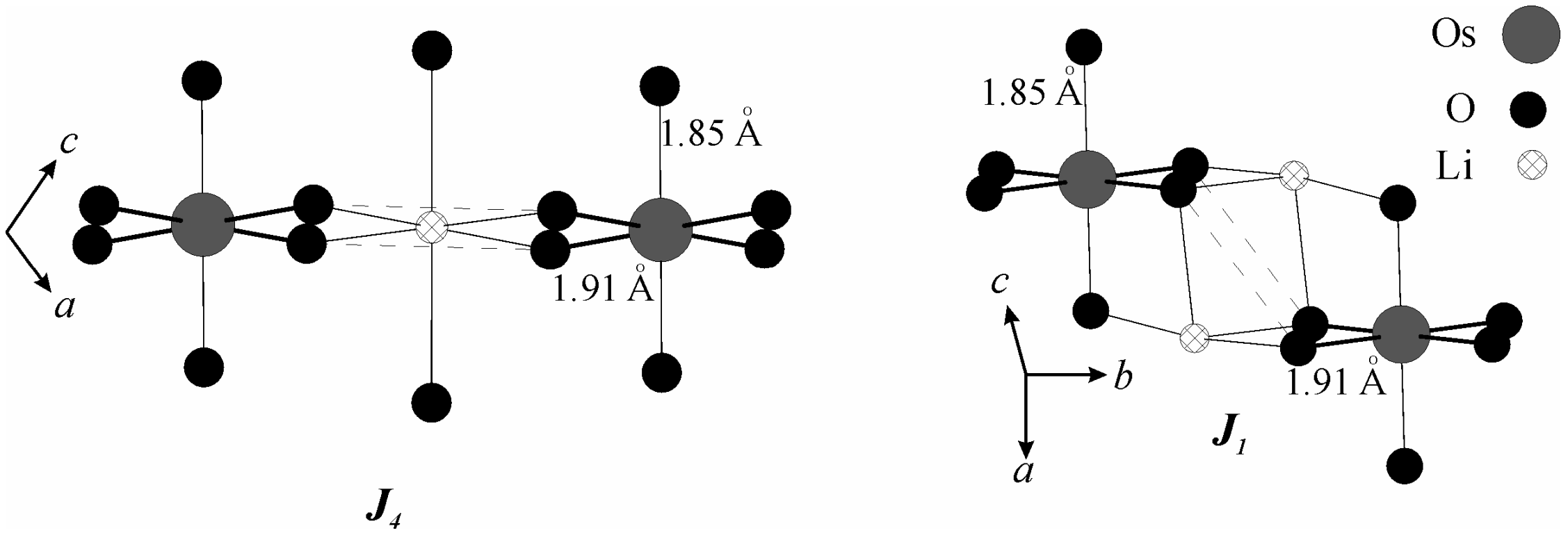

**Figure 7:** The detailed $J_1$ and $J_4$ super-super-exchange pathways. The long and short *M*–O bonds are shown by the thick and thin lines, respectively. The O–O interactions are denoted by the dashed lines. The four oxygen, two osmium and the lithium atom, which are involved in the $J_4$ pathway (left) are coplanar. The $J_1$ pathway (right) does not fulfill this condition.

From the alternative approach, when the contributions of $d_{xy}$, $d_{xz}$ and $d_{yz}$ to the exchange interactions are equally probable and the influence of the tetragonal compression is minimized, a very different result is found. In fact $J_2$, $J_3$ and $J_4$ which form triangles in the *ac* plane are the three dominant interactions now. In such cases geometrical magnetic frustration is expected.

**Table IX.** <($\Delta e$)$^2$> for the various exchange pathways in both $Li_5OsO_6$ and $Li_4MgReO_6$ calculated on the spin dimer model, assuming that the occupation of the $d_{xy}$, $d_{xz}$ and $d_{yz}$ orbitals are equally probable.

| | $Li_5OsO_6$ | | | | | | $Li_4MgReO_6$ | | | | | |
|---|---|---|---|---|---|---|---|---|---|---|---|---|
| | <Δe> (meV)$^2$ | | | Rel. | | | <Δe> (meV)$^2$ | | | Rel. | | |
| x [a] | 0 | 0.05 | 0.10 | 0 | 0.05 | 0.10 | 0 | 0.05 | 0.10 | 0 | 0.05 | 0.10 |
| $J_1$ | 20 | 48 | 189 | 0.03 | 0.03 | 0.04 | 51 | 79 | 428 | 0.06 | 0.03 | 0.08 |
| $J_2$ | 710 | 1800 | 4128 | **1** | **0.96** | **0.95** | 893 | 2320 | 5374 | **1** | **1** | **1** |
| $J_3$ | 310 | 1881 | 4334 | **0.44** | **1** | **1** | 723 | 1890 | 4492 | **0.81** | **0.81** | **0.84** |
| $J_4$ | 420 | 888 | 1865 | **0.59** | **0.47** | **0.43** | 626 | 1343 | 2648 | **0.70** | **0.58** | **0.49** |
| $J_5$ | 18 | 42 | 78 | 0.03 | 0.04 | 0.02 | 27 | 69 | 133 | 0.03 | 0.03 | 0.02 |

a: x is the modification factor for the diffuse STO exponent of O *2p* in the form of $(1\text{-}x)\times \zeta'_i$

The energy levels of the *6d* orbitals obtained from the extended Hückel molecular orbital calculations for $[OsO_6]^{5-}$ and $[ReO_6]^{6-}$are shown in Table X.

**Table X.** The molecular orbital energies of *6d* states for $[OsO_6]^{5-}$ and $[ReO_6]^{6-}$ octahedra.

| | 25(eV) | 26(eV) | 27(eV) | 28(eV) | 29(eV) |
|---|---|---|---|---|---|
| $[OsO_6]^{5-}$ | -10.003 | -9.814 | -9.809 | -2.208 | -0.965 |
| $[ReO_6]^{6-}$ | -10.545 | -10.464 | -10.435 | -2.627 | -2.041 |

It is evident that the energy splitting between the lowest lying $d_{xy}$ state to the next state for $[OsO_6]^{5-}$is larger than that of $[ReO_6]^{6-}$, 0.189 eV compared to 0.081 eV. This indicates that the Os-based compound exhibits a stronger *Jahn-Teller* distortion and therefore is best described based on the first approach without magnetic frustration. On the other hand, $Li_4MgReO_6$ does not show strong tendency towards tetragonal compression and results in more closely spaced energy states. Thus the results of Table X are more relevant here, supporting the geometrical magnetic frustration and is in good accordance with the previously reported experimental investigations.[11]

The low temperature neutron diffraction data for $Li_5OsO_6$ did not show any extra peaks due to long range magnetic ordering. This is disappointing but not unusual for such dilute systems (one magnetic atom out of twelve atoms) with $S$ = ½. Therefore, it is impossible to propose a magnetic structure for $Li_5OsO_6$.

Note that the calculated transfer energies, $<\Delta e^2>$, are significantly larger for all $J$'s for $Li_4MgReO_6$ relative to $Li_5OsO_6$. In addition size of $Os^{7+}$ is smaller than that of $Re^{6+}$, which results in larger Hubbard U for $Os^{7+}$. Accordingly, the spin exchange interactions for $Li_4MgReO_6$ are expected to be larger than those of $Li_5OsO_6$. This observation is semi-quantitatively consistent with the much larger, negative $\theta$ value for the Re-based system.

## IV. SUMMARY AND CONCLUSIONS

The crystal structure of the ordered rock salt type $Li_5OsO_6$ was reinvestigated, using room temperature powder neutron diffraction data. This compound crystallizes in $C2/m$ space group with the lattice parameters $a$ = 5.0472(1) Å, $b$ = 8.7827(2) Å, $c$ = 5.0079(1) Å, $\beta$ = 109.777(2)°. There is no mixing between $Os^{7+}$ and $Li^+$ in the crystal structure. The material shows AFM long range order below 40K and a frustration index f ~ 1. Nonetheless, evidence for short range, low dimensional, AFM correlations is found from both susceptibility and heat capacity data when entropy removal is taken into account. This behavior is different from that of the isoelectronic, isostructural compound, $Li_4MgReO_6$, which exhibits spin glass behavior below 12K and f ~ 14.

Spin dimer analysis was performed to understand the origin of this remarkable contrast. The results were consistent with a frustrated triangular lattice model for the Re-based compound while a stronger tetragonal compression in $Li_5OsO_6$ encourages low dimensional magnetic correlations rather than geometrical magnetic frustration.

**ACKNOWLEDGMENTS**

We thank Paul Dube for his assistance in collecting magnetic susceptibility and heat capacity data. We appreciate fruitful discussions with Craig Bridges and Mario Bieringer. J. E. G. acknowledges the Natural Sciences and Engineering Research Council of Canada for the financial support of this work.